\documentclass[%
12pt,
reprint,
amsfonts,amsmath,amssymb,
aps,
prb,
]{revtex4-1}

\usepackage{mathrsfs}
\usepackage{amssymb}
\usepackage{amsbsy}
\usepackage{color}
\usepackage{cancel}
\usepackage[tight]{subfigure}
\usepackage{graphicx}
\usepackage[tight]{subfigure}
\usepackage{dcolumn}
\usepackage{bm}
\usepackage[colorlinks,linkcolor={blue},citecolor={blue},urlcolor={blue}]{hyperref}
\usepackage{marginnote}
\usepackage{modiagram}
\usepackage{upgreek}
\usepackage{relsize}

\begin{document}

\title{Intrinsically high thermoelectric figure of merit of half-Heusler ZrRuTe}

\author{Sonu Prasad Keshri}\email{sonuprasadkeshri16@iisertvm.ac.in} 
\author{Amal Medhi}\email{amedhi@iisertvm.ac.in} 
\affiliation{Indian Institute of Science Education and Research Thiruvananthapuram, 
Kerala 695551, India}

\begin{abstract}
The electronic structure and thermoelectric properties of ZrRuTe-based Half-Heusler
compounds are studied using density functional theory (DFT) and Boltzmann transport formalism.
Based on rigorous computations of electron relaxation time $\tau$ considering electron-phonon 
interactions and lattice thermal conductivity $\kappa_l$ considering phonon-phonon interactions,
we find ZrRuTe to be an intrinsically good thermoelectric material. It has a high power
factor of $\sim 2\times 10^{-3}$ W/m-K$^{2}$ and low $\kappa_l\sim 10$ W/m-K at 800 K. The thermoelectric figure of merit $ZT \sim 0.13$ at 800 K is higher than 
similar other compounds. We have also studied the properties of the material as a function of doping and find the thermoelectric properties to be substantially enhanced for $p$-doped ZrRuTe with the $ZT$ value raised to $\sim 0.2$ at this temperature. 
The electronic, thermodynamic, and transport properties of the material are thoroughly studied and discussed. 
\end{abstract}

\maketitle

\section{Introduction}
Thermoelectric (TE) devices hold great promise in technological applications in today's 
world as it can act as clean sources of energy converting waste heat into 
electricity as well as robust devices in refrigeration technology to transfer heat
from a cold to hot reservoir\cite{Sundar_2014,Bell_2008}. 
The chief drawback of these devices has been their low efficiency. 
Current research in this area are intensely focused on finding new
materials with high thermoelectric efficiency\cite{Sundar_2014,Bell_2008,Poon_2019,Poon_2015}, 
a quantity which is measured by the figure of merit ($ZT$)\cite{Poon_2015},
\begin{equation}
ZT = \frac{S^2 \sigma T}{\kappa_e+\kappa_l} 
\end{equation}
The quantities $S$, $\sigma$, and $\kappa_e$ are Seebeck coefficient, electrical 
conductivity, and electronic thermal conductivity, respectively, and are related 
to the electronic contribution to the transport phenomena. $T$ is the absolute 
temperature, and $\kappa_l$ is phonon contribution to thermal conductivity called 
lattice thermal conductivity. The problem in getting a high $ZT$ value is 
all the above quantities are interrelated and can't be optimized independently. 
Various routes have been explored over the years to enhance $ZT$, such as reducing lattice thermal conductivity by increasing
phonon scattering by using the technique of 
doping\cite{Routes_2018,Zhao141_2016,Hu_2014,Tritt_2017,Poon_2018}, nano-structuring\cite{Newtrend_2017,Routes_2018,Poon_2019}, alloying\cite{Poon_2015,Poon_2018} without significantly changing the electronic contribution to transport properties. 
Also several different classes of materials such as Skutterudites\cite{Skutteru_1996}, PbTe\cite{PbTe_2016}, SnSe\cite{Zhao141_2016,SnSe_2016}, $\mathrm{Bi_{2}Se_{3}}$, $\mathrm{Bi_{2}Te_{3}}$\cite{Mishra_1997} has been identified where $ZT$ values are inherently higher.
The Half-Heusler (HH) compounds have been one of the primary target material
in this regard with great potential especially for high temperatures thermoelectric
applications \cite{Zhu_HH,Shuo_2013,He_2016,Yang_2008}, given its several favorable 
properties such as good thermal and mechanical
stability, absence of toxic elements, eco-friendly properties etc.

The HH compounds have the general composition of XYZ, where X, Y are transition metals, 
and Z is a main group element. 
In the band structure of HH, a strong $d$ hybridization between the X, Y transition 
elements and $d$-$p$ hybridization between Y, Z elements occur, all of which are 
responsible for opening a bandgap at Fermi level. 
Localized $d$ orbitals give flat band edges causing low mobility 
and high Seebeck coefficient whereas $s$, $p$ orbitals give rise to high electrical
conductivity. The carrier concentration, the degeneracy of bands, the density of states 
and the effective mass etc.\ all form the controlling factors 
in determining the thermoelectric properties of these materials.

There are several compounds in the HH group that have already been studied and 
reported to have good thermoelectric properties. This includes 
$n$-type MNiSn-based compounds, $p$-type MCoSb-based
compounds (M=Ti, Zr, Hf)\cite{Sekimoto_2007}, $p$-type FeNbSb-based 
compounds\cite{Chenguang_Zhu_2015}, $p$-type NbInSn\cite{Hohl_1998} etc. The compound
$\mathrm{FeNb_{1-x}Hf_xSb}$\cite{Shi_PRB_2017,Chenguang_2015} is found to be one of the best HH compounds
with a $ZT\sim 1.5$ at 1200 K. Another widely studied HH alloy is $\mathrm{Zr_xHf_{1-x}NiSn_ySb_{1-y}}$\cite{Shen_2001,Poon_2019,Tritt_2017,Poon_2015,Poon_2018}, which is also reported to have $ZT>1$
for certain combinations of $x$ and $y$. The $ZT$ values of the HH compounds are generally high
because of its higher power factors. However, their one demerit has been the high thermal conductivity 
$\kappa$. For example, $\kappa$ 
(in units of W/m-K) of $\mathrm{FeNb_{0.88}Hf_{0.12}Sb}$
is $4$ at 1200 K\cite{Shi_PRB_2017} and that of 
$\mathrm{Zr_{0.5}Hf_{0.5}NiSn_{0.99}Sb_{0.01}}$
is $6$ at 800 K\cite{Shen_2001}. 
These values are relatively much higher than typical
thermoelectric material like $\mathrm{Bi_2Te_{2.3}Se_{0.7}}$ for which $\kappa$ 
is 1.2 at 500 K\cite{Hu_2014} and hole-doped SnSe with $\kappa$ of 0.55 at
773K\cite{Zhao141_2016}. However, the HH family of compounds offers enormous possibility to
search for the best thermoelectric material as it comprises of thousands of different 
possible compounds, many of which are still unexplored.

The ZrRuTe is one such compound that has not been studied so far. Like other semiconducting
HH materials, this compound also has a valence electron count of 18 and hence expected to be a
semiconductor. It comprises of elements with $4d$ electrons. Hence electrical conductivity and 
the power factor are expected to be higher compared to $3d$ systems. At the same time, 
due to the heavier masses of the elements, the lattice thermal conductivity is expected to be 
lower, making it a possible high-efficiency thermoelectric material. 
In this work, we study the equilibrium and thermoelectric transport properties of 
ZrRuTe in detail using {\em ab-initio} methods. We first determine the stable crystal structure from phonon dispersion calculated using density functional perturbation theory (DFPT). 
The electronic structure is determined using density functional theory (DFT) and the thermoelectric transport coefficients are computed within the semiclassical Boltzmann transport formalism. An accurate estimation of the transport coefficients is a challenging task. Especially the electron relaxation time 
$\tau$ is 
very expensive to compute from ab-initio theory, and often its value is deduced in an ad-hoc basis
from experimental data for similar compounds. Here we estimate $\tau$ rigorously for ZrRuTe
by considering electron-phonon interactions computed using maximally localized Wannier wavefunctions. 
Also, lattice thermal 
conductivity $\kappa_l$ is determined considering 
phonon-phonon interactions involving both two and three phonon processes. From the calculations, 
we find ZrRuTe to be an intrinsically good thermoelectric material with a high power
factor of $\sim 2\times 10^{-3}$ W/m-K$^{2}$ comparable to similar other like FeNbSb and ZrNiSn based HH compounds\cite{Shi_PRB_2017,Chenguang_2015,Shen_2001}.
Undoped ZrRuTe have a high thermoelectric figure of merit $ZT$ $\sim 0.134$ at 800 K, which much higher than undoped TiCoSb, ZrCoSb, and HfCoSb at 973 K\citep{Takeyuki_2005}.
 We have also studied the properties of the material as a function of doping and find the thermoelectric properties to be substantially enhanced for $p$-doped $\mathrm{ZrTc_xRu_{1-x}Te}$
($x=0.125)$ with the $ZT$ value raised to $\sim 0.2$ at this temperature. 
Overall, we find the material to be a promising system for thermoelectric applications.
The electronic, thermodynamic and transport properties of the material are thoroughly studied and discussed.

The rest of the paper is organized as follows. In section~\ref{sec:comp_details}, we 
give the details of the computational technique and methodology used to calculate the electronic, thermodynamic, and transport properties of the material. In subsections
 ~\ref{Crystal structure}-\ref{phonon dispersion}, we discuss the structural properties and phonon dispersion. In subsections ~\ref{electronic structure}-\ref{electronic transport coeff}, the electronic structure and the electrical transport coefficients like electrical conductivity, Seebeck coefficient, electronic thermal conductivity, and power factor as a function of temperature and doping are discussed. The lattice thermal conductivity of ZrRuTe is briefly analyzed and discussed in the subsection ~\ref{lattice thermal conductivity}.

\section{Computational details}
\label{sec:comp_details}
The half-Heusler compounds that we study here are undoped ZrRuTe, hole-doped 
ZrTc$_x$Ru$_{1-x}$Te and electron-doped ZrRu$_{1-x}$Rh$_{x}$Te ($x=0.125$).
The electronic structure is determined using density functional theory (DFT) as 
implemented in the QUANTUM ESPRESSO package\cite{QE_2017}. 
First, we take supercell of 24-atoms and performed geometry optimization by variable cell relaxation and then fixed volume relaxation for all three compounds. 
The SCF calculations were done for supercells on a Monkhorst-Pack k-mesh of size $8\times8\times8$  
and an energy cutoff of 50 Ry for ZrRuTe and $\mathrm{ZrTc_{x}Ru_{1-x}Te}$. 
The electron-doped system converged with larger $k$-mesh of $12\times12\times12$ with an energy cutoff of 60 Ry. We did the band structure calculations along high symmetry paths with the same parameters as for SCF. The density of states calculation is done with the large k-mesh size of $20\times20\times20$ for ZrRuTe and $\mathrm{ZrTc_{x}Ru_{1-x}Te}$ compounds and with an increased k-point mesh of $24\times24\times24$ for $\mathrm{ZrRu_{1-x}Rh_{x}Te}$ compound with tetrahedra smearing. The convergence threshold of $10^{-8}$ Ry is taken for the self-consistency for all three compounds. 
We used Perdew-Zunger (LDA) exchange correlational functional with 
Rappe Rabbe Kaxiras Joannopoulos (ultrasoft) pseudopotential.\\
The phonon calculation needs both DFT and DFPT calculation which we did on $4\times4\times4$ q-mesh with a threshold value of $10^{-14}$ for 
self-consistency for phonon calculation, followed by the SCF calculation on 
Monkhorst-Pack k-point mesh of $2\times2\times2$ with `conv-thr' of $10^{-12}$ 
and energy cutoff of 40 Ry. 
We took a convergence threshold of $10^{-10}$ (a.u)
on the total energy and $10^{-8}$ (a.u) on the total force for ionic minimization.
The Fourier interpolation of the band structure is done to calculate the electronic transport coefficients within the semiclassical BTE formalism in constant relaxation time approximation (RTA) implemented in the
BoltzTrap\cite{Boltztrap} package. The lattice thermal conductivity is 
calculated using the ShengBTE\cite{Shengbte_2014} package. 
The ShengBTE code needs three input files - the $2^{nd}$ order interatomic force constant (IFC), the $3^{rd}$ order IFC, and geometry and internal input, which includes information about interpolation. The $2^{nd}$ order interatomic force constant is calculated 
using DFPT technique\cite{Baroni_phonon_2001,Tang_2011} by calculating the 
dynamical matrix. The third-order anharmonic IFC is calculated up to 4 nearest 
neighbors using the finite difference method\cite{Stokes_2008,Shengbte_2014}. Here, a 
supercell of size $4\times4\times4$ having 192 atoms is used to 
create displacement. This step is computationally very demanding requiring
handling of several hundreds of the computer-generated files.   
The lattice thermal conductivity calculation is done on $10\times10\times10$ q-point 
grid with a scale broad (gaussian broadening) of 0.01.\\
 A crucial quantity in the study of transport properties is the electron
relaxation time $\tau$. Accurate estimation of $\tau$ is very hard as 
theoretically, one has to make various approximations and numerical computation
is very expensive. The BoltzTraP package gives the
electronic transport coefficients only in units of $\tau$. 
Here, we make an {\it ab initio} calculation of $\tau$ using the EPW 
package\cite{S.Ponce_2016}, which computes the electron-phonon scattering rates 
using DFPT and maximally localized Wannier functions (MLWF)\cite{Marzari_2012}.
Using the EPW code, we compute the electron-phonon scattering rate 
$\Gamma_{i,\vec{k}}^{FM}$ and from the imaginary part of the 
Fan-Migdal electron self-energy as 
$\Gamma_{i,\vec{k}}^{FM}=(\frac{2}{{\hbar}})\mathrm{Im}\sum_{i,\vec{k}}^{FM}$ 
and then relaxation time as $\tau_{i,\vec{k}}=(\Gamma_{i,\vec{k}}^{FM})^{-1}$.

\section{results and analysis}
\label{sec:results and analysis}
\subsection{Crystal structure}
\label{Crystal structure}
Half-Heuslers have F-43m space group of XYZ composition where elements 
X and Z form $\mathrm{XY_4}$ and $\mathrm{ZY_4}$ tetrahedra structure 
in the nearest neighbor coordination (Fig.~\ref{fig:HH_Structure}). 
Each Y atom lies at the center of an $\mathrm{X_4Z_4}$ cube 
forming $\mathrm{YX_4}$ and $\mathrm{YZ_4}$ tetrahedra.
Thus all the three positions 4b (1/2, 1/2, 1/2), 4c (1/4, 1/4, 1/4) and 4a (0, 0, 0) of XYZ have $\mathrm{T_d}$ symmetry\cite{Djung_2000,Tanja_2011}. 
It may be mentioned that both formula XYZ and YXZ are equivalently used in 
literature. Still, the interchange of position of atoms in actual crystal structure matters
as they may show different semiconducting, semimetallic, or metallic states.
We optimized both possible structures ZrRuTe and RuZrTe and chose the most stable structure ZrRuTe, which we are reporting to be a semiconductor. We find
the distance between nearest-neighbor atoms to be $\frac{\sqrt{3}}{4}a$ and 
that between next-nearest neighbor atom to be $\sim$ 15.47\% larger. The lattice 
parameter ($a$) is calculated to be 6.2107 {\AA} using LDA. 
A very tiny change in the lattice parameter is seen on the substitution of Ru atoms with Tc or Rh atoms. This change is positive of 0.0036 {\AA} for Tc substitution and negative of 0.0042 {\AA} for Rh substitution.
\begin{small}
\begin{figure}[htbp]
 \centering
 \subfigure[]{\label{fig:HH crystal structure}
 \includegraphics[width=0.8\columnwidth]{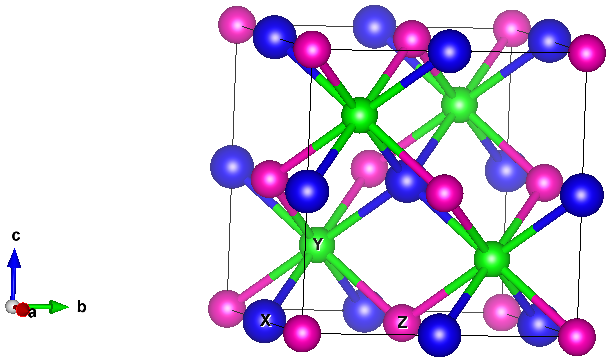}}
 \subfigure[]{\label{fig:Zr}
  \includegraphics[width=0.3\columnwidth]{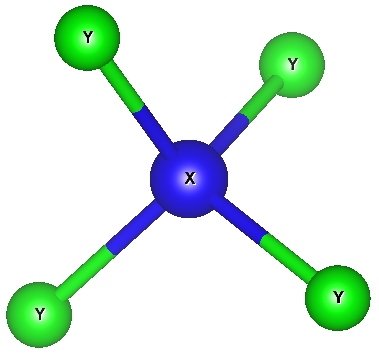}}
 \subfigure[]{\label{fig:Ru} 
  \includegraphics[width=0.3\columnwidth]{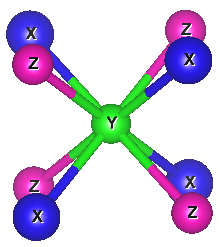}}
 \subfigure[]{\label{fig:Te} 
  \includegraphics[width=0.3\columnwidth]{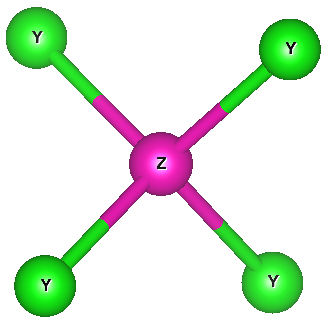}}
  \caption{\subref{fig:HH crystal structure} Crystal structure of HH compounds XYZ.
  \subref{fig:Zr}-\subref{fig:Te} Tetrahedral coordination of X (Zr), Y (Ru), 
  and Z (Te) atoms.}
 \label{fig:HH_Structure}
\end{figure}
\end{small}
\begin{figure}[htbp]
 \centering
 \includegraphics[width=0.7\columnwidth]{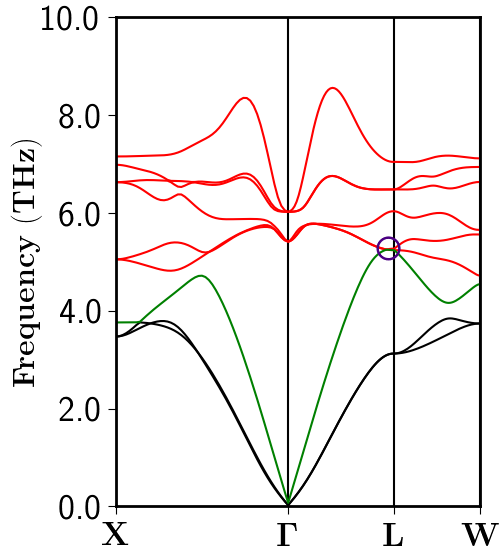}
  \caption{Phonon dispersion relation of ZrRuTe.
There are nine phonon modes resulting from three atom per unit cell. 
The black lines indicate two transverse acoustic (TA), green line indicates
one longitudinal acoustic (LA) and red lines indicate six optical (O) modes.
The circle shows the LA-TO band touching.}
\label{ph_dis}
\end{figure}

\subsection{Phonon dispersion and structural stability}
\label{phonon dispersion}
To ascertain the thermodynamical stability, we investigate the
phonon spectrum of the compound and indeed find no negative frequency mode in it,
as shown in Fig.~\ref{ph_dis}. The maximum frequency obtained 
is 8.56 THz for the optical mode, which is comparable to ZrCoSb\cite{Shiomi_2011}. This implies that the bonding is as strong as that in ZrCoSb.
At $\mathrm{\Gamma}$ point, there are three degenerate regions, one acoustic and two optical regions of lower and higher frequencies. In all three regions, the
transverse modes are doubly degenerate. There exist small LO-TO splittings
of 0.057 $\mathrm{cm^{-1}}$ in the high-frequency region and of 0.0569 
$\mathrm{cm^{-1}}$ in low-frequency region due to non-analytic nature of the
dynamical matrix in the limit $\mathrm{q}\rightarrow 0$ resulting from
the long-range nature of Coulomb interaction in polar materials.
Remarkably this splitting at high frequency is much smaller than the value 
of 50 $\mathrm{cm^{-1}}$ reported for ZrCoSb\cite{Shiomi_2011}.
The TA modes and high frequency TO modes remain degenerate along $\Gamma-$L, but
the two low frequency TO modes get split by 0.97 $\mathrm{cm^{-1}}$ at L.
On the other hand, the double degeneracy of high frequency TO modes is broken
along $\Gamma-$X. 
Down in the lower energy region near L, a 
band touching is seen between the LA and TO modes corresponding
to frequencies 174.6688 $\mathrm{cm^{-1}}$, 174.5569 $\mathrm{cm^{-1}}$ 
and 174.5461 $\mathrm{cm^{-1}}$ indicated by the circle in Fig.~\ref{ph_dis}. 
This behavior plays a significant role in determining the thermal transport 
properties\cite{Ouyang_2018} of the ZrRuTe compound.

\subsection{Electronic structure analysis}
\label{electronic structure}
\begin{tiny}
\begin{figure}
  \centering

  \begin{MOdiagram}[lines={gray,thin},AO-width =20pt,names]
    \small
    \AO[metal-4d-1](40 pt){s}{1.38;}
    \AO[metal-4d-2](40 pt){s}{1.40;}
    \AO[metal-4d-3](40 pt){s}{1.48;}
    \AO[metal-4d-4](40 pt){s}{1.50;}
    \AO[metal-4d-5](40 pt){s}{1.52;}
    \AO[metal-5s](40 pt)  {s}{3.10;}
    \AO[metal-5p-1](40 pt){s}{4.25;}
    \AO[metal-5p-2](40 pt){s}{4.27;}
    \AO[metal-5p-3](40 pt){s}{4.29;}

    \AO[ligand-1](180 pt){s}{1.80;}
    \AO[ligand-2](180 pt){s}{1.60;}
    \AO[ligand-3](180 pt){s}{2.10;}
    \AO[ligand-5](180 pt){s}{2.30;}

    \AO[complex-a1]   (110 pt){s}{-0.60;}
    \AO[complex-t'2] (110 pt){s}{0.75;}
    \AO[complex-e] (110 pt){s}{0.98;}
    \AO[complex-t2] (110 pt){s}{1.94;}
    \AO[complex-e*]  (110 pt){s}{2.90;}
    \AO[complex-t*2] (110 pt){s}{4.00;}
    \AO[complex-t'*2] (110 pt){s}{4.70;}
    \AO[complex-a*1] (110 pt){s}{5.30;}

    \connect
      {
        metal-4d-4     & complex-t'2   ,
        metal-4d-4     & complex-t2    ,
        metal-4d-4     & complex-t*2   ,
        metal-4d-4     & complex-t'*2  ,
        metal-4d-2     & complex-e     ,
        metal-4d-2     & complex-e*    ,
        metal-5s       & complex-a1    ,
        metal-5s       & complex-a*1   ,
        metal-5p-2     & complex-t'2   ,
        metal-5p-2     & complex-t2    ,
        metal-5p-2     & complex-t*2   ,
        metal-5p-2     & complex-t'*2  ,        
        complex-t'2    & ligand-1      ,
        complex-t2     & ligand-1      ,
        complex-t*2    & ligand-1      ,
        complex-t'*2   & ligand-1      ,
        complex-a1     & ligand-2      ,
        complex-a*1    & ligand-2      ,
        complex-e      & ligand-3      ,
        complex-e*     & ligand-3      ,
        complex-t'2    & ligand-5      ,
        complex-t2     & ligand-5      ,
        complex-t*2    & ligand-5      ,
        complex-t'*2   & ligand-5      ,                
        }

    \node[left] at (metal-4d-3.west) {$4\mathrm{d}$};
    \node[below] at (metal-4d-1.west) {$\mathrm{e}$};
    \node[above] at (metal-4d-5.west) {$\mathrm{t}_{2}$};
    \node[above] at (metal-5s.west) {$\mathrm{a}_{1}$};
    \node[above] at (metal-5p-3.west) {$\mathrm{t}_{2}$};
    \node[left] at (metal-5s.west)   {$5\mathrm{s}$};
    \node[left] at (metal-5p-1.west) {$5\mathrm{p}$};

    \node[below] at (complex-a1)   {$\mathrm{a}_{1}$};
    \node[below] at (complex-t'2) {$\mathrm{t}^{'}_{2}$};
    \node[above] at (complex-e)  {$\mathrm{e}$};
    \node[below] at (complex-t2) {$\mathrm{t}^{}_{2}$};
    \node[above] at (complex-e*)  {$\mathrm{e}^{*}$};
    \node[above] at (complex-t*2){$\mathrm{t}^{*}_{2}$};
    \node[above] at (complex-t'*2){$\mathrm{t}^{'*}_{2}$};
    \node[above] at (complex-a*1){$\mathrm{a}^{*}_{1}$};
    
    \node[right] at (ligand-1.east) {\footnotesize $\mathrm{t}_{2}$};
    \node[right] at (ligand-2.east) {\footnotesize $\mathrm{a}_{1}$};
    \node[right] at (ligand-3.east) {\footnotesize $\mathrm{e}$};
    \node[right] at (ligand-5.east) {\footnotesize $\mathrm{t}_{2}$};
    \node[right] at (200 pt, 48 pt) {$\upsigma$};
    \node[right] at (200 pt, 65 pt) {$\pi$};

    \draw[orange, <->] (complex-t2.center) -- (complex-e*.center)
      node[midway,left] {band gap} ;

    \node at (  40 pt, -40 pt) {$\mathrm{Ru^{4-}}$};
    \node at ( 110 pt, -40 pt) {$\mathrm{(RuTe)^{4-}}$};
    \node at (180 pt, -40 pt) {$\mathrm{Te}$};
    \EnergyAxis[title=E,head=stealth]
  \end{MOdiagram}
  \caption{Most probable MO energy-level scheme for regular tetrahedral structure with  $\upsigma$ and $\pi$-donor legand.}
  \label{fig:MO}
\end{figure}
\end{tiny}

\begin{figure}[htb]%
\centering
\includegraphics[width=0.80\columnwidth]{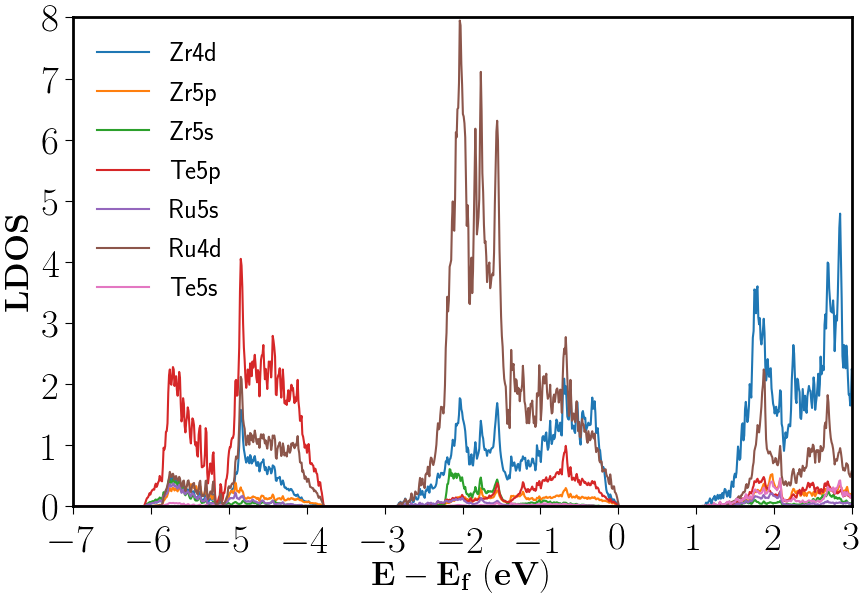}%
\caption{Local density of states (LDOS) of undoped ZrRuTe. The zero of energy is taken at the Fermi level.}%
\label{fig:ldos_lda}%
\end{figure}

In HH compounds, all the three atoms have $\mathrm{T_{d}}$ symmetry 
in the first coordination sphere and octahedral symmetry in the next.
Here we shall focus on only nearest-neighbor interactions and neglect the 
next nearest interactions, which because of its multipole electrostatic nature,
falls very sharply with distance.  
The tetrahedral (point charge crystal field) interactions between
Zr and Ru, and hybridization between the $s$ and $d$ orbitals lead to the
formation of approximately closed shelled $\mathrm{Zr^{4+}}$ ($\mathrm{4d^{0}5s^{0}}$) and $\mathrm{Ru^{4-}}$ ($\mathrm{4d^{10}5s^{2}}$) ions.
This happens because of the electronegativity of Zr atom (1.33) is much smaller than
that of the Ru atom (2.2) which allows the shared electrons to surround the Ru
atom more compared to the Zr atom.
Now there is a strong coordinate-covalent interaction between the $\mathrm{Ru^{4-}}$ ion and four Te atoms, thus reducing the three atom system to effectively a
two atom ($\mathrm{Ru^{4-}}$ ion-Te atom) system. The schematics of
interactions of the symmetry allowed orbitals of $\mathrm{Ru^{4-}}$ and Te are shown in Fig.~\ref{fig:MO}. The molecular orbitals are filled with electrons 
according to Hund's rule and we get a closed-shell MO electronic configuration 
($\mathrm{a^{2}_{1}t^{'6}_{2}e^{4}t^{6}_{2}}$) with 18-valence electrons 
for $\mathrm{(RuTe)^{4-}}$ which form a zinc blende structure\cite{Tanja_2011}.

\begin{figure*}[!htb]%
\centering
\subfigure[]{%
\label{fig:0dope}%
\includegraphics[width=0.70\columnwidth]{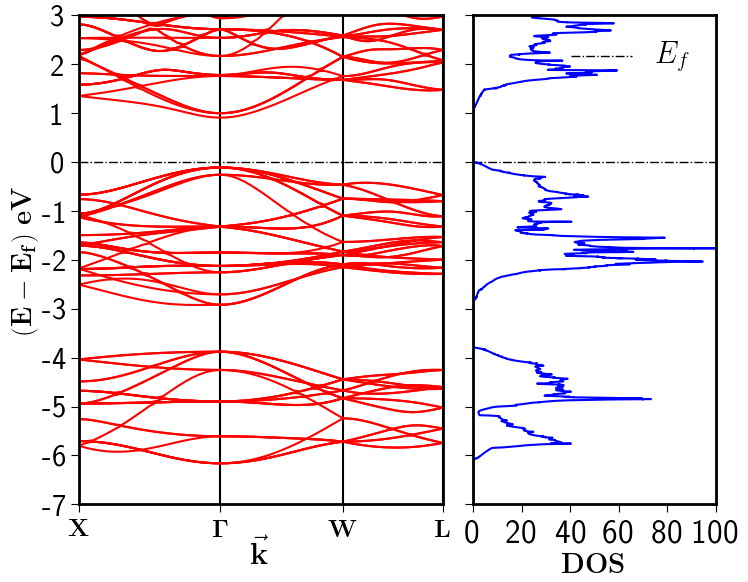}}%
\subfigure[]{%
\label{fig:ndope}%
\includegraphics[width=0.70\columnwidth]{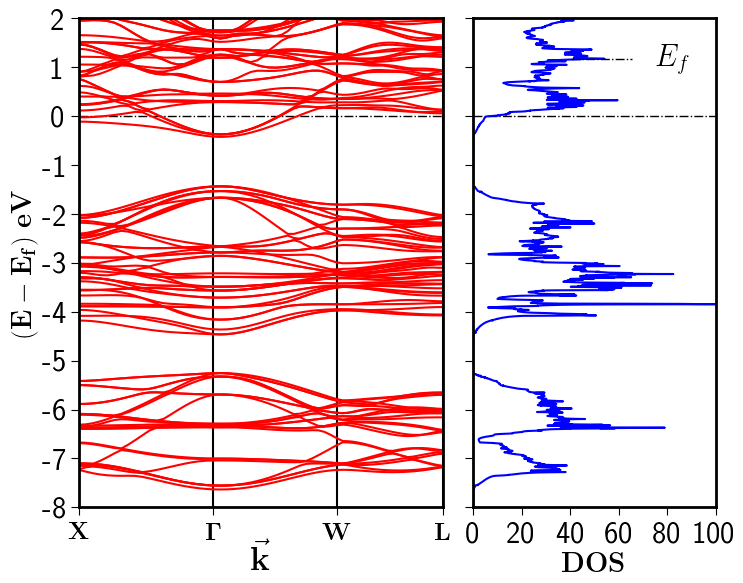}}%
\subfigure[]{%
\label{fig:pdope}%
\includegraphics[width=0.70\columnwidth]{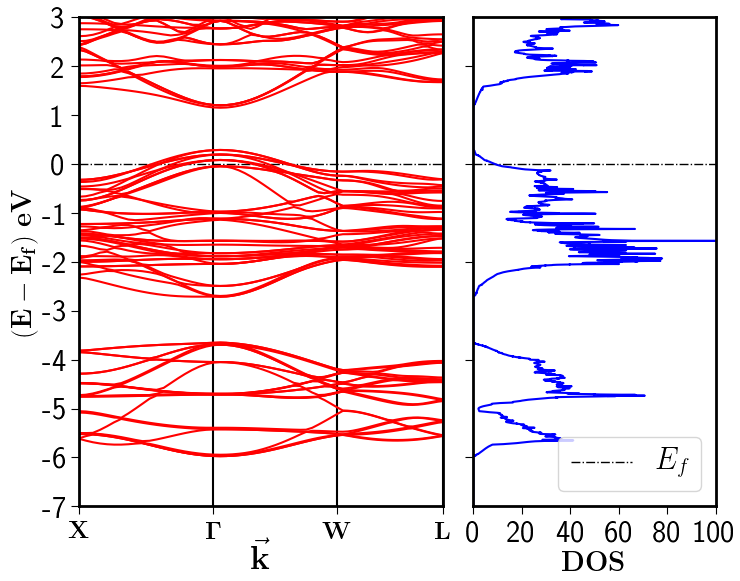}}%
\caption{Band structure and density of states (DOS) of
(a) undoped (b) $n$-doped, and (c) $p$-doped ZrRuTe. The zero of energy is set at the Fermi level.}%
\label{fig:band_structure}
\end{figure*}

The contribution to the density of states (DOS) from orbitals of each atom is shown in Fig.~\ref{fig:ldos_lda} and the band structure in Fig.~\ref{fig:0dope}. 
The DOS around $-$12 eV to $-$13 eV (not shown in the figure) mainly comes from 
Te-$5s$ orbital of symmetry $\mathrm{a_{1}}$. In the energy range of $-6.1$ to 
$-3.8$ eV, a substantial contribution comes from Te-5$p$ and lesser 
from Ru-$4d$ and Zr-$4d$ orbitals of $t'_2$ and $e$ symmetries.
Despite the strong hybridization of orbitals, an energy gap of $\sim 1$ eV 
from around $-$3.8 to $-$2.8 eV manifests due to the tetrahedral e-$\mathrm{t_2}$ splitting. The valence band ($-2.8$ to 0 eV) which comes mainly from 
Ru-$4d$ orbitals also contain significant contributions from
Zr-$4d$ and Te-$5p$ orbitals of $t_2$ symmetry. Near the valence band edge, 
the quadratic like dispersing bands comes from the $p$-orbitals of
the Te atoms. Whereas the $d$-orbitals of Ru and Zr dominate flat band dispersion around the energy of $-1.5$ to $-2.3$ eV. 
In Figs.~\ref{fig:ndope} and \ref{fig:pdope}, we show the band structures of 
the system doped with $n$ and $p$-type impurities, respectively.
The doping lifts some of the degeneracies at high symmetry points due to the
lowering of the crystal symmetry by the presence of impurity atoms.

\subsection{Electrical transport coefficients}
\label{electronic transport coeff}
In this section, we discuss the electrical transport properties of the material.
We use the Boltzmann transport equation (BTE) formalism in constant relaxation time approximation to calculate the transport coefficients. Within the formalism, 
the Seebeck coefficient $S_{\alpha \gamma}$, the electrical conductivity 
$\sigma_{\alpha \gamma}$, and the electronic thermal conductivity 
$\kappa^{e}_{\alpha \gamma}$ are given by\cite{Boltztrap}
\begin{small}
\begin{align}
S_{\alpha \gamma} (T,\mu) = &\frac{1}{eT\Omega \sigma_{\alpha \gamma}(T,\mu)}{\mathlarger{\int}} \bar{\sigma}_{\alpha \gamma}(\varepsilon)(\mu-\varepsilon) \bigg[\frac{\partial f_{\mu}(T,\varepsilon)}{\partial \varepsilon}\bigg] d\varepsilon \\
 \sigma_{\alpha\gamma} (T,\mu) = &\frac{1}{\Omega}{\mathlarger{\int}} \bar{\sigma}_{\alpha \gamma}(\varepsilon) \bigg[-\frac{\partial f_{\mu}(T,\varepsilon)}{\partial \varepsilon}\bigg] d\varepsilon \\
 \kappa_{\alpha\gamma} (T,\mu) = &\frac{1}{e^{2}T\Omega}{\mathlarger{\int}} \bar{\sigma}_{\alpha\gamma}(\varepsilon)(\varepsilon - \mu)^{2}\bigg[-\frac{\partial f_{\mu}(T,\varepsilon)}{\partial \varepsilon} \bigg]d\varepsilon
\label{eq:ele_thermalconductivity_tensor}
\end{align}
\end{small}\\
where $\alpha$, $\gamma$ are tensor indices. $\Omega$, $\mu$, and $f$ are respectively
the volume of the unit cell, chemical potential, and the Fermi-Dirac 
distribution function in local equilibrium. The transport distribution 
function tensor is defined as
\begin{equation}
\bar{\sigma}_{\alpha\gamma}(\varepsilon) = \frac{e^2}{N} {\mathlarger{\sum}}_{i,\vec{k}} \tau\times \vartheta _{\alpha}(i,\vec{k})\vartheta _{\gamma}(i,\vec{k})\frac{\delta (\varepsilon-\varepsilon_{i,\vec{k}})}{d\varepsilon}
\end{equation}
where $\vec{k}$ is the wave vector, and $i$ is the band index. N is the number of $\vec{k}$ points sampled, $\tau$ is the carrier relaxation time and $\vartheta_{\gamma}$ is the carrier group velocity along $\gamma$ direction,
\begin{equation}
\vartheta_{\gamma}(i,\vec{k}) = \frac{1}{\hbar}\frac{\partial \varepsilon _{i,\vec{k}}}{\partial {k_{\gamma}}}
\label{eq:group_velocity}
\end{equation} 
All the electrical transport coefficients are calculated in terms of relaxation time $\tau$. Thus to know the absolute values one must calculate the relaxation time as well, accurate computation of which is a highly non-trivial task. To estimate $\tau$ one often resorts to a simplistic approach like deformation potential approximation\cite{Fang_2017,Fang_2017} or use experimental results for similar compounds to infer its value\cite{Sekimoto_2007}. In general, the relaxation time depends on details of k-points and band index in the first Brillouin zone, therefore care must be taken in the sampling of reciprocal space.
For the undoped system, the electron-phonon interaction is the dominating source of scattering mechanism.
Here we estimate $\tau$ and mobility for undoped ZrRuTe rigorously by calculating electron-phonon interactions using the Wannier wavefunctions implemented in the EPW code\cite{S.Ponce_2016}.
In this scheme, electron mobility is calculated in self-energy relaxation time approximation using\cite{Samuel_2018}
\begin{align}
{\mu}_{e,\alpha\gamma} =& \frac{-e}{n_e\Omega}\mathlarger{\mathlarger{\sum}}_{i\in CB}\int \frac{d\vec{k}}{\Omega_{BZ}}\Bigg[\frac{\partial f_{\mu}(T,\varepsilon_{i,\vec{k}})}{\partial \varepsilon_{i,\vec{k}}}\Bigg]\tau_{i,\vec{k}}\times \nonumber \\ &\vartheta _{\alpha}(i,\vec{k})\vartheta _{\gamma}(i,\vec{k})
\label{eq:mobility}
\end{align}
where $i$ is the band index, $n_e$ is the number density of electrons, 
$\Omega_{BZ}$ is the volume of the first Brillouin zone.
$\varepsilon_{i,\vec{k}}$ is the single electron eigenvalue
and $\tau_{i,\vec{k}}=(\frac{2}{\hbar}\mathrm{Im}\sum_{i,\vec{k}}^{\mathrm{FM}})^{-1}$\cite{Samuel_2018,Giustino_2017} 
is the carrier relaxation time. The relaxation rate is given by
\begin{align}\small
 \frac{1}{\tau_{i,\vec{k}}}=&\frac{2\pi}{\hbar}\sum_{j\nu}\int\frac{d\vec{q}}{\Omega_{BZ}}|g_{ji\nu}(\vec{k},\vec{q})|^{2}[(f_{i,\vec{k}}+n_{\vec{q}\nu})\times \nonumber\\
& \delta(\varepsilon_{i\vec{k}}-\varepsilon_{j\vec{k}+\vec{q}}-\hbar\omega_{\vec{q}\nu})
+(1-f_{i,\vec{k}}+n_{\vec{q}\nu})\times \nonumber\\
&\delta(\varepsilon_{i\vec{k}}-\varepsilon_{j\vec{k}+\vec{q}}+\hbar\omega_{\vec{q}\nu})]
 \label{eq:tau2}
\end{align}
where $\omega_{\vec{q}\nu}$ is the phonon frequency, $\vec{q}$ is the phonon wave vector, $\nu$ is the branch index, $n_{\vec{q}\nu}$ is the Bose-Einstein distribution function, $\sum_{i,\vec{k}}^{\mathrm{FM}}$ is the Fan-Migdal electron self-energy 
and $g_{ji\nu}(\vec{k},\vec{q})$ is the probability amplitude for scattering 
from an initial electronic state $|i\vec{k}\rangle$ to final state 
$|j\vec{k}+\vec{q}\rangle$ via a phonon $|\nu\vec{q}\rangle$. 
A similar equation as Eq.~(\ref{eq:mobility}) can be obtained for 
holes as the charge carriers.

The $\tau_{i,\vec{k}}$ calculated above is averaged over the states near the Fermi level to obtain
$\bar{\tau}$. Average mobility is obtained from averaging the absolute value of mobility along three axes X, Y and Z. 
The calculated values of $\bar{\mu}$ and $\bar{\tau}$ for undoped ZrRuTe are given in 
Table~\ref{table:table3}. 
\begin{table}
\centering
\begin{tabular}{ |p{1.1cm}|p{1.4cm}|p{1.4cm}|p{1.4cm}|p{1.2cm}|p{1.2cm}| }
 \hline
 \multicolumn{6}{|c|}{Mobility ($\mathrm{cm^{2}/Vs}$) and relaxation time (fs)} \\
 \hline
 T (K)&$E_f$ (eV)&$\bar{\mu_e}$&$\bar{\mu_h}$ &$\bar{\tau_{h}}$&$\bar{\tau_{e}}$ \\
 \hline
 300 & 16.16 & 33.67& 34.73&-&- \\
 \hline
 800 & 16.16 & 3.62& 7.34&1.23&0.59\\
 \hline
\end{tabular}
\caption{Average relaxation time $(\bar{\tau})$ and average mobility $(\bar{\mu})$ due to electron-phonon scattering
for undoped ZrRuTe.}
\label{table:table3}
\end{table}
\begin{figure*}[!htb]
 \centering
 \small
 \subfigure[]{\label{fig:Sigma}
 \includegraphics[width=0.48\columnwidth]{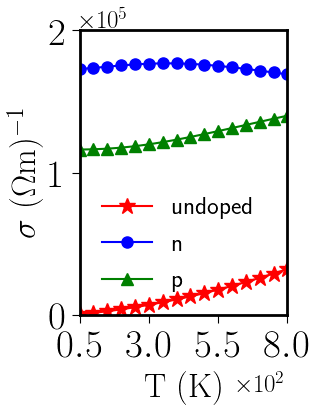}}
 \subfigure[]{\label{fig:Seebeck}
  \includegraphics[width=0.5\columnwidth]{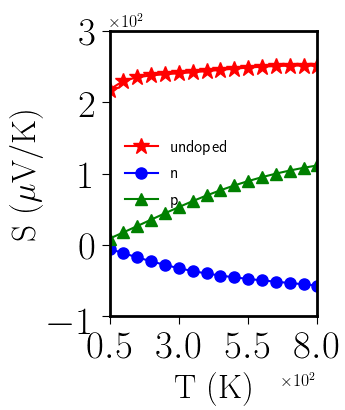}}
 \subfigure[]{\label{fig:PF} 
  \includegraphics[width=0.5\columnwidth]{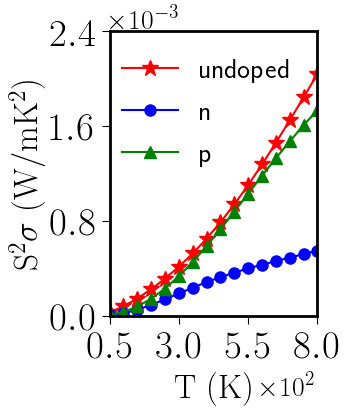}}
 \subfigure[]{\label{fig:kappa} 
  \includegraphics[width=0.48\columnwidth]{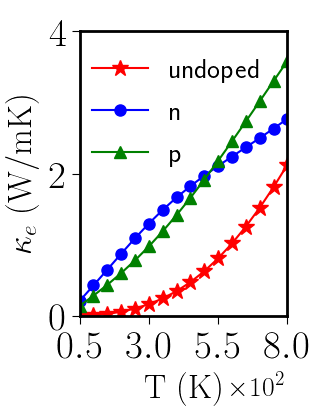}}
   \subfigure[]{\label{fig:sigma_mu}
 \includegraphics[width=0.48\columnwidth]{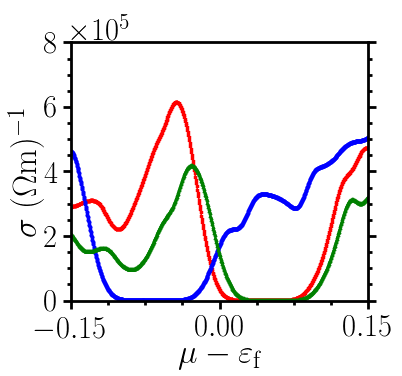}}
  \subfigure[]{\label{fig:Seebeck_mu}
 \includegraphics[width=0.49\columnwidth]{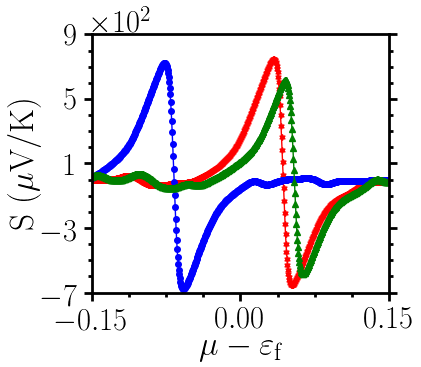}}
  \subfigure[]{\label{fig:pg_mu}
 \includegraphics[width=0.48\columnwidth]{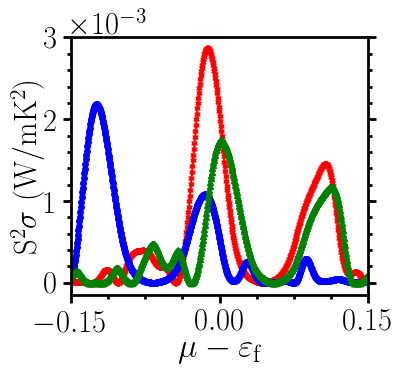}}
  \subfigure[]{\label{fig:ke_mu}
 \includegraphics[width=0.48\columnwidth]{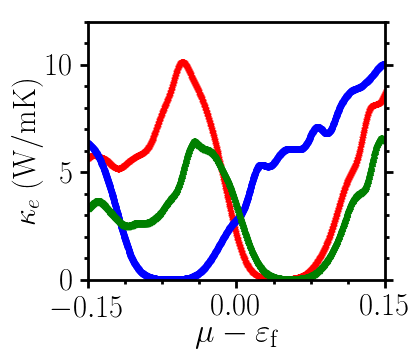}}
\caption{Transport coefficients of ZrRuTe. In all the figures, the colors red, blue, and green
correspond to data for undoped, $n$-doped, and $p$-doped compounds, respectively.
In \subref{fig:Sigma} to \subref{fig:kappa}, we show electrical
conductivity $\sigma$, thermoelectric power $S$, power factor $S^2\sigma$, and 
electronic thermal conductivity $\kappa_e$ as functions of temperature $T$ (50 K to 800 K).
In \subref{fig:sigma_mu} to \subref{fig:ke_mu}, we show $\sigma$,
$S$, $S^2\sigma$, and $\kappa_e$ as functions of chemical potential $\mu$ (in Ry) at 800 K.}
\label{fig:electric_transport}
\end{figure*}
The data shows that total mobility ($\bar{\mu_{h}}+\bar{\mu_{e}}$) 
decreases with increasing temperature which happens due to increase in electron-phonon 
scatterings at higher temperatures.
The effective masses of electrons and holes are given by $\mu_{e,h} = e\tau_{e,h}/m^{*}_{e,h}$ 
are estimated to be $m^{*}_{e}=0.286 m_{e}$ and $m^{*}_{h}=0.278 m_{e}$ at 800 K, respectively.
For the doped system, in principle, one also needs to take into account impurity scatterings to 
deterimine the total relaxation time. However in general for fully ionized impurities, 
the relaxation time due to impurity scattering time $\tau_{im}$ goes as $T^{\frac{3}{2}}$ whereas electron-phonon relaxation time $\tau$ goes as $T^{\frac{-3}{2}}$ \cite{J.M_Ziman}. Therefore at high temperatures, the impurity scatterings can be safely ignored. For the electron-phonon part, we take $\tau$ for the doped system to be upper limit estimate of $\tau$ for the undoped ZrRuTe as the computations for the doped system is prohibitively expensive.

Next, we proceed to calculate the other transport coefficients. 
The results for electrical conductivity $\sigma$ are shown in Fig.~\ref{fig:Sigma}. 
For undoped ZrRuTe, the $\sigma$ increases with temperature confirming the semiconducting behavior, 
which is evident from the band structure. The temperature dependence of conductivity follows
$\sigma  = aT^{1.53}$ with $a=1.12$. The hole-doped system is metallic at zero temperature as
the Fermi level lies in the valence band. Indeed this metallic behavior persists up to 52 K as we show 
in Fig.~\ref{fig:psigma_10_50} below. At higher temperatures, electrons jump 
into the conduction band overcoming the band-gap of 0.8 eV. Hence the semiconducting behavior 
beyond this temperature, as seen in Fig.~\ref{fig:Sigma}. The variation of $\sigma$ with temperature
in the range 52 to 800 K follows the 
relation  $\sigma  = aT^{\frac{5}{3}} + \sigma_{0}$ 
with $a= 3.52\times 10^{-1}$ and $\sigma_{0}=1.15\times 10^{5}$.
The $n$-doped compounds have different electrical behavior as compared to the
other two. A metallic behavior at low temperature is expected from the band structure as the Fermi
level falls in the conduction band. However Fig.~\ref{fig:Sigma} shows a semiconducting trend 
for $T$ up to $370$ K and metallic beyond.
This could be due to the presence of localized impurity states at the Fermi level.
This behavior is experimentally seen for similar compounds like ZrCoSb and HfCoSb\cite{Xia_2000,Takeyuki_2005,Sekimoto_2007}.
In the metallic region, the conductivity obeys the empirical relation,
$\sigma  \simeq aT^{\frac{5}{2}} + \sigma_{0}$
with $a= -5.33\times 10^{-4}$ and $\sigma_{0}=1.78\times 10^{5}$ $\mathrm{(\Omega m)^{-1}}$.
It is evident that both n-type and p-type compounds have roughly four times more electrical conductivity at 800 K than undoped one. 
\begin{figure}
 \subfigure[]{\label{fig:psigma_10_50}
 \includegraphics[width=0.48\columnwidth]{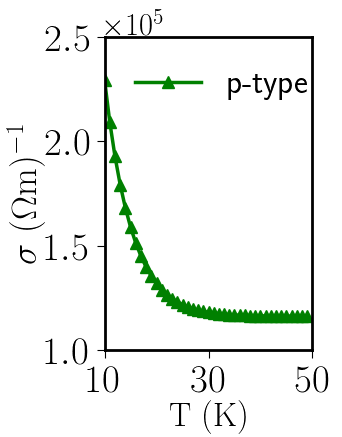}}
  \subfigure[]{\label{fig:Cmol_e}
 \includegraphics[width=0.48\columnwidth]{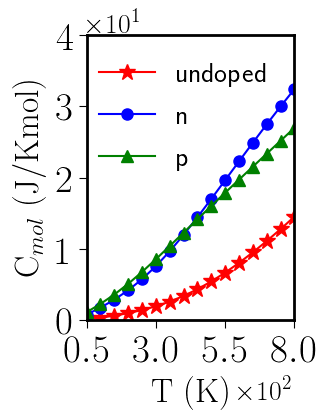}}
\caption{
\subref{fig:psigma_10_50} Electrical conductivity of p-doped compound in temperature range 10 to 50 K, showing metallic bahavior. \subref{fig:Cmol_e} electronic specific heat per mole in temperature range 50 to 800 K.}
\end{figure}

The Seebeck coefficient $S$ of a semiconductor is closely related to DOS at the Fermi level, and can be controlled by tunning the band structure. The Fig.~\ref{fig:Seebeck} shows $S$ for undoped and doped compounds as a function of $T$. The drop in $S$ with doping as clear from the figure is due to 
increase in DOS at the Fermi level which falls in either valence or conduction band for these compounds. The value of $S$ for undoped ZrRuTe at $800$ K is ~$\sim$ 250 $\mu$V/K, which is roughly twice the value for TiCoSb\cite{Takeyuki_2005} and nearly equal to the value for doped ZrCoSb at 1000 K\cite{Sekimoto_2007}. This value also compares well with the estimates for similar HH compounds\cite{Yustandnyk_2010,Yoshisato_2010}. Our calculation show, the values of $S$ are $\sim 110$  and $\sim -53$ (in units of $\mu$V/K) at 800 K for the $p$-doped and
the $n$-doped compounds, respectively. The positive and the negative values of $S$ are signatures of
the dominating charge carriers being holes and electrons, respectively. The magnitude of $S$ in all the
cases increases monotonically with temperature in the range 50 K to 800 K. The power factor 
$S^2\sigma$ increases with $T$ albeit at different rates for the undoped, $n$-doped and $p$-doped
compounds as shown in Fig.~\ref{fig:PF} . AT 800 K, the values of $S^2\sigma$ for the undoped, $n$-doped and $p$-doped compounds are $2.04$, $0.47$ 
and $1.77$, respectively, in units of $10^{-3}$ W/m-K$^{2}$.
These values are comparable to the highest earlier reported power factors of 
doped FeNbSb and ZrNiSn based compounds\cite{Shi_PRB_2017,Chenguang_2015,Shen_2001}.

The transfer of heat by the charge carriers and phonons in a thermoelectric material is an 
undesirable factor and it has been one of the major challenges to achieving low thermal conductivity in
various thermoelectric materials. Doping a materials with impurity atoms reduces the lattice 
part of the thermal conductivity but increases the electronic contribution $\kappa_e$ due to 
increased density of states (DOS) at the Fermi level. Our calculated values of $\kappa_e$ at a function
of $T$ are shown in Fig.~\ref{fig:kappa}. At 800 K, $\kappa_e$ for the undoped, $n$-type and $p$-type compounds are found to be $\sim$ 2.0, 2.56 and 3.47 W/m-K, respectively. The quantity can also be related to the molar specific heat by 
$\kappa_{e} = \frac{1}{3}C_{mol}\bar{v}^{2}\tau$. The quantity $C_{mol}$ can be
calculated using the equation
\begin{align}
 C_{mol}  = &{\mathlarger{\int}} n(\varepsilon)(\varepsilon - \mu)\bigg[\frac{\partial f_{\mu}(T,\varepsilon)}{\partial \varepsilon} \bigg]d\varepsilon
\end{align}
Our calculated values of $C_{mol}$ are shown in Fig.~\ref{fig:Cmol_e} as a function of temperature. 
The higher value of the specific heat for the doped systems is clearly due to the enhanced DOS 
$n(\varepsilon)$, as mentioned. 
Further analysis of the results shows that the material does not follow the Wiedemann-Franz law, 
$\kappa_{e}=L \sigma T$, $L$ being the Lorentz number. For instance, $\kappa_{e}/\sigma T$ for the
undoped ZrRuTe is $1.26L$ at $T=50$ K, while it is $1.48L$ at $T=800$ K.
We discuss the lattice thermal conductivity in the next section and show that it dominants over the electronic part of the thermal conductivity in undoped ZrRuTe.

\subsection{Lattice thermal conductivity}
\label{lattice thermal conductivity}
\begin{figure*}[htb]
 \centering
 \small
  \subfigure[]{\label{fig:latt_conv}
 \includegraphics[width=0.67\columnwidth]{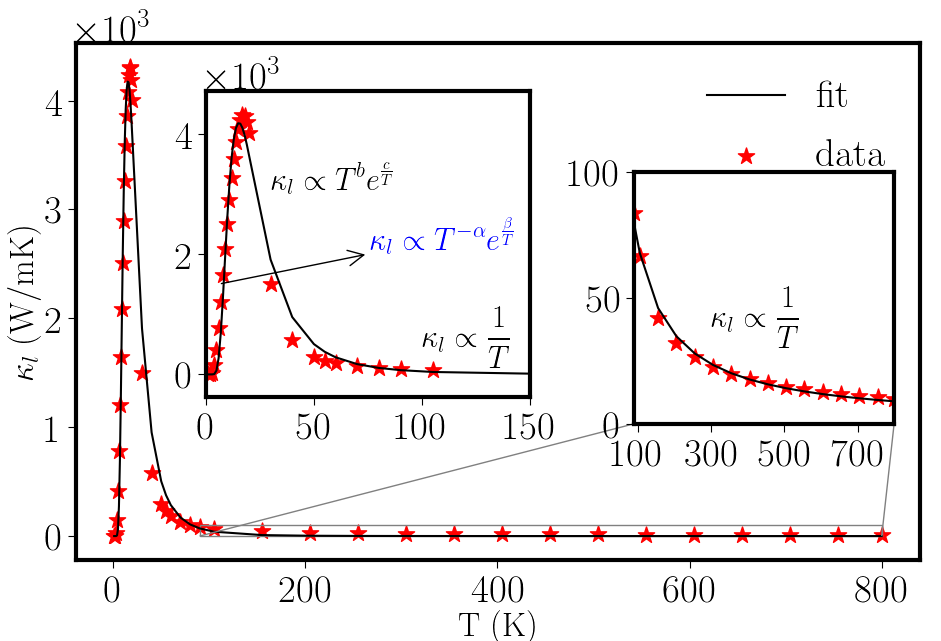}}
 \subfigure[]{\label{fig:Cv_latt}
  \includegraphics[width=0.6\columnwidth]{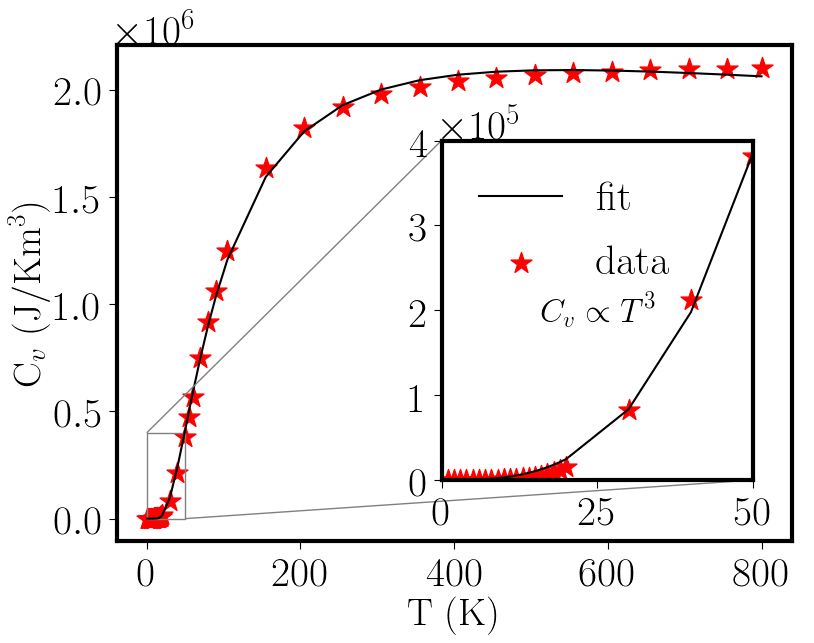}}
  \caption{\subref{fig:latt_conv} Lattice thermal conductivity of $\kappa_l$ undoped ZrRuTe in temperature range 1 to 800 K. The insets show how $\kappa_l$ varies with $T$ in different temperature ranges. \subref{fig:Cv_latt} Lattice specific heat per unit volume in temperature range 1 to 800 K for undoped ZrRuTe as function of temperature. The inset shows Debye law at low temperatures.}
\end{figure*}

The lattice thermal conductivity $\kappa_l$ is computed using the ShengBTE package\cite{Shengbte_2014}
which takes into account phonon-phonon interactions, phonon scatterings by impurities such as isotopes,
and also scatterings due to finite boundaries. The method solves the resulting linearized Boltzmann transport equation (LBTE) iteratively. The zeroth order of iteration is equivalent 
to the relaxation time approximation (RTA) in which $\kappa_l$ is given by
\begin{align} 
\kappa^{\alpha\beta}_{l}=\frac{\hbar^{2}}{k_{B}T^{2}\Omega N}\sum_{\lambda}f_{0}(f_{0}+1)\tau_{\lambda}\omega^{2}_{\lambda}\vartheta^{\alpha}_{\lambda}\vartheta^{\beta}_{\lambda}
\end{align}
where $\vartheta_{\lambda}$, $\omega_{\lambda}$, and $\tau_{\lambda}$ are group velocity, angular
 velocity, and relaxation time respectively of phonons of mode $\lambda$, and $f_{0}$ is the 
 equilibrium Bose-Einstein distribution function.
The scheme requires computations of second and third-order interatomic force constants (IFC) of the lattice, which is computationally very expensive. Here
we carry out the calculations for undoped ZrRuTe and the results are shown in Fig.~\ref{fig:latt_conv}.
At 800 K, $\kappa_l$ for the system obtained within RTA and after full iteration give values
9.83 and 9.97 W/m-K, respectively which are roughly equal to the values
for well-known half-Heusler compounds ZrCoSb and TiCoSb at 700 K\citep{Takeyuki_2005}. 
The temperature variation of $\kappa_l$ shows distinct behaviors in three different temperature 
ranges. In the range 90-800 K, $\kappa_l$ shows the $\frac{1}{T}$ variation due to Umklapp processes.
This variation can also be understood from the kinetic formula $\kappa_{l} = \frac{1}{3}C_{v}\bar{v}l_{ph}$. At high temperatures, the specific heat $C_v$ and average speed $\bar{v}$ are constants whereas
mean free path $l_{ph}$ varies as $1/T$. Indeed we estimated the Debye temperature $\Theta_D$ for
the system within harmonic approximation and found it to be $\sim 90$ K.
In the lower temperature range of 17 to 90 K, $\kappa_l$ can be fitted to 
a $T^{b}\exp(\frac{c}{T})$ variation, $b$ and $c$ being the fitting parameters. At very low
temperatures from 1 to 17 K, $\kappa_l$ varies as 
$\kappa_{l} \sim \gamma T^{-\alpha}e^{\frac{-\beta}{T}}$ where 
$\gamma \simeq 2.7\times 10^{5}$, $\alpha \simeq 0.93$ and $\beta \simeq 25.25$.
This deviation from the Debye law at $T\ll\Theta_D$ is due to the fact that the boundary effects 
become important at such low temperatures.
 
For doped compounds, computations of the $2^{nd}$ and $3^{rd}$-order IFCs with supercells 
becomes prohibitively expensive, and we do not undertake the task. Instead we take $\kappa_l$
for the doped systems to be the one-third of the value obtained for undoped ZrRuTe. This is 
based on results for similar other compounds, e.g.~$\mathrm{Fe(V_{0.6}Nb_{0.4})_{1-x}Ti_{x}Sb}$\cite{Fu_2014}, $\mathrm{ZrCoBi_{1-x}Sn_{x}}$\cite{ZhuHangtian_2018} and $\mathrm{Hf_{1-x}Zr_{x}NiSn_{1-y}Sb_{y}}$\cite{YU20092757} where $\kappa_l$ has been found to be reduced nearly by a factor of 3 
by around 15\% dopings.

\subsection{Thermoelectric figure of merit}
\begin{figure}
\includegraphics[width=0.6\columnwidth]{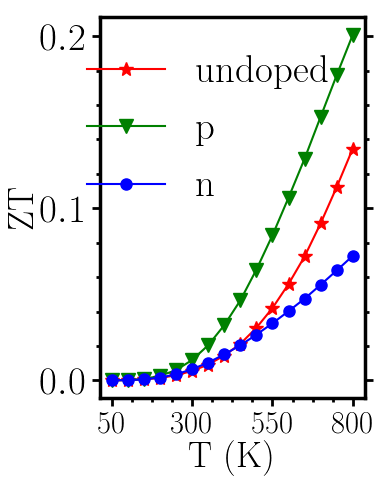}
\caption{The dimentionless figure of merit ($ZT$) as function of temperature.}
\label{fig:ZT}
\end{figure}

Having computed all the transport coefficients, 
here we estimate the figure of merit $ZT$ for the thermoelectric efficiency 
of the material. The results obtained for undoped ZrRuTe and well as 
its $p$-type and $n$-type doped counterparts are shown as a function
of temperature in Fig.~\ref{fig:ZT}. From the results, overall, we 
find that ZrRuTe is an intrinsically good thermoelectric material.
The $ZT$ value for the undoped system is $\sim 0.13$ at 800 K which is 
significantly higher than the experimentally reported values for 
HH compounds like TiCoSb, ZrCoSb and HfCoSb with $ZT$ values 0.01, 0.02, 
and 0.027, respectively at 973 K\citep{Takeyuki_2005}. This high 
value is because of the higher power factor of the material, which in turn
results from its favorable band structure properties. The power factor
remains almost unaffected by $p$-type doping, whereas it gets 
reduced by $n$-type doping, as shown in Fig.~\ref{fig:PF}. However, the simultaneous
reduction of lattice thermal conductivity $\kappa_l$ by doping means that
$ZT$ for $p$-type ZrRuTe gets significantly enhanced, as seen in 
Fig.~\ref{fig:ZT}. Its value of $\sim 0.2$ at $800$ K is one of the 
highest values among the $p$-type half-Heusler (HH) compounds. 
It is worthily to mention that here we have considered a doping 
concentration $x$ of $12.5$ percent. There is further scope for 
improvement of the $ZT$ value as a function of $x$ which is not explored here.
Thus given its intrinsically high power factor, $p$-doped ZrRuTe-based 
compounds seem to be a promising thermoelectric material worthwhile for further experimental studies.

\section*{Conclusion}
To conclude, we have made detail first-principles calculations of 
thermoelectric properties of half-Heusler compounds based on ZrRuTe and 
discussed the results thoroughly. The structural stability of the compounds is 
confirmed from the calculation of phonon dispersions. The electronic band structure 
is investigated in detail and analyzed thoroughly. Transport coefficients are calculated by rigorous 
estimation of the electron relaxation time $\tau$ from electron-phonon interactions.
We find that because of the favorable band structure, the power factor of ZrRuTe is higher
with a value $\sim 2\times 10^{-3}$ W/m-K$^2$ for the undoped system, compared to 
similar other compounds like TiCoSb, ZrCoSb and HfCoSb. The lattice thermal conductivity 
of value $\sim 10$ W/m-K for undoped ZrRuTe is comparable to that of some of the well known good half-Heusler thermoelectric materials like XCoSb (X=Ti,Zr,Hf). Overall the thermoelectric figure of 
merit is found to be higher for the $p$-doped system with a $ZT \sim 0.2$ at 800 K, which is one of the highest value reported for $p$-type half-Heusler compounds. Thus ZrRuTe is a promising thermoelectric material worthwhile for further experimental studies.

\bibliography{references}
\end{document}